\documentclass[aps,prd,twocolumn,showpacs,floatfix,preprintnumbers,nofootinbib
]{revtex4-1}

\usepackage{graphicx}
\usepackage{hyperref}
\usepackage{color}
\usepackage{natbib}
\usepackage{multirow}
\usepackage{amsmath}
\usepackage{amssymb}
\usepackage{epsfig}
\usepackage{textcomp}
\usepackage{color}
\usepackage{ulem}

\def\theta{\vartheta}

\newcommand{\be}{\begin{equation}}
\newcommand{\ee}{\end{equation}}
\newcommand{\ba}{\begin{eqnarray}}
\newcommand{\ea}{\end{eqnarray}}
\newcommand{\lsim}   {\mathrel{\mathop{\kern 0pt \rlap
  {\raise.2ex\hbox{$<$}}}
  \lower.9ex\hbox{\kern-.190em $\sim$}}}
\newcommand{\gsim}   {\mathrel{\mathop{\kern 0pt \rlap
  {\raise.2ex\hbox{$>$}}}
  \lower.9ex\hbox{\kern-.190em $\sim$}}}

\begin{document}

\title{A new evaluation of the antiproton production cross section for cosmic ray studies}
\author{Mattia di Mauro$^{1,2}$, Fiorenza Donato$^{1}$, Andreas Goudelis$^{2}$, Pasquale Dario Serpico$^{2}$}
\affiliation{$^{1}$Physics Department, Torino University, and  Istituto Nazionale di Fisica Nucleare, Sezione di Torino, via Giuria 1, 10125 Torino, Italy}
\affiliation{$^{2}$LAPTh, Univ. de Savoie, CNRS, B.P.110, Annecy-le-Vieux F-74941, France}

\begin{abstract}
Theoretical predictions for the cosmic antiproton spectrum currently fall short of the corresponding experimental level of accuracy. Among 
the main sources of uncertainty are the antiproton production cross sections in cosmic ray inelastic interactions.
We analyse existing data on antiproton production in $pp$ scattering, including for the first time 
the measurements performed by the NA49 Collaboration. 
We compute the antiproton spectrum finding that in the energy range where data are available (antiproton energies of about 
4-550 GeV) 
different approaches lead to almost equivalent results,  with an uncertainty of 10-20\%. 
Extrapolations outside this region lead to different estimates, with the uncertainties reaching the 50\%
level around $1$ TeV, degrading the diagnostic power of the antiproton channel at those energies. 
We also comment on the uncertainties in the antiproton production source term coming from nuclei heavier
than protons and from neutrons produced in $pp$ scatterings, and point out the need 
for dedicated experimental campaigns for all processes involving antiproton production in collisions of light nuclei. 
\end{abstract}

\preprint{LAPTH-051/14}


\maketitle

\section{Introduction}\label{Intro}
Cosmic ray (CR) antiprotons are a remarkable diagnostic tool for astroparticle physics. The bulk of the measured flux is certainly consistent with
a purely secondary origin in CR collisions onto interstellar medium gas, but additional {\it primary} components are not excluded,  
either of astrophysical origin (see for instance~\cite{Blasi:2009bd})
or of exotic nature, such as dark matter annihilation or decay~\cite{Salati:2010rc}. 
At very least, antiprotons provide a consistency check for the current understanding of galactic CR modeling and can narrow down propagation parameters 
(see e.g. \cite{Maurin:2002ua,Evoli:2008dv,Donato:2008jk}).

This tool is however only as sharp as the uncertainties entering the background (i.e. the secondary component) and signal 
(i.e. the primary component) computations are robust. 
Statistical and systematic errors reported by the PAMELA collaboration~\cite{Adriani:2010rc} are already at the 10\% level 
up to the 10 GeV scale, below the theoretical error. In a short time, AMS-02~\cite{AMS02web} is expected to provide significantly higher precision, 
calling for a reassessment of the theoretical predictions.

The contribution of different processes to the ${\bar p}$ secondary yield has been studied in the past, see e.g.~\cite{Moskalenko:1998id,Donato:2001ms,Moskalenko:2001ya}.
In~\cite{Donato:2001ms}, for instance, the uncertainties on the production cross sections were estimated to be $\sim 25\%$, and
already identified as the limiting factor in theoretical predictions (see also~\cite{Moskalenko:2001ya} for similar considerations).
In practice, nuclei heavier than protons and helium only contribute at a few percent level (see e.g.~\cite{Moskalenko:1998id}), 
thus playing a very marginal role, either as projectiles or targets, in the antiproton production. 
Reactions involving helium (p-He, He-p, He-He) represent a sizable 
fraction of the total yield, easily reaching $\sim 50\%$ at low energies \cite{Donato:2001ms}.

While for processes involving helium nuclei no data is available, the situation is different for the
proton-proton case, where there are several experimental studies. The latest re-evaluation of the antiproton production yield in 
$pp$ collsions was reported in~\cite{Duperray:2003bd}, while the Tan \& Ng parameterization~\cite{Tan:1982nc} is still largely used,
despite being more than 30 years old. The reason is that, until recently, the available dataset was limited to data only collected in the
sixties and seventies. In the last decade, however, two more experimental datasets have become available: the BRAHMS data~\cite{Arsene:2007jd}   
and---more important for the energies of interest for AMS-02 applications---the NA49  results collected at the 
CERN Super Proton Synchrotron (SPS)~\cite{Anticic:2009wd}. 

Given the importance of these nuclear data for new measurements in astroparticle physics, it seems thus timely to re-evaluate the antiproton production 
cross section in $pp$ collisions in light of this newly available information. In this paper we engage ourselves in this task, 
in order to provide the community with a  parametrization for the inclusive antiproton production 
cross section as well as with a reliable assessment of the corresponding uncertainties that should be taken into account 
in CR studies.

The outline of the paper goes as follows: 
in section \ref{sec:formdatamethod} we set up the relevant formalism, present the experimental data that will be used subsequently 
and describe the analysis methods. 
In section \ref{sec:results} we present our results. We begin by validating our analysis framework by reproducing existing results
in the literature and then move on to evaluate the inclusive antiproton production cross section, first relying solely on the novel
NA49 data and then on the full set of available measurements.
In section \ref{sec:therest} we briefly comment on the impact  of other contributions entering 
the secondary antiproton source term, namely of antineutrons and helium nuclei.
Finally, in section \ref{sec:discussionconclusions} we discuss our key results and present our conclusions.
Two appendices follow, where we present some---standard but useful---kinematics used in our analysis as well as a new evaluation
of the total, elastic and inelastic $pp$ scattering cross sections that we performed for the energy range of interest to our work.
\section{Framework, data and methods}\label{sec:formdatamethod}

\subsection{Theoretical framework}\label{sec:formalism}

CR protons interact with the interstellar medium (ISM) and may produce secondary antiprotons.  Different channels are involved, with the
dominant one being the  CR proton flux collisions with the target hydrogen gas ($pp$). 
The corresponding source term is  the convolution 
of the antiproton production cross section $ \frac{d\sigma_{p \, p \rightarrow \bar{p}}}{dE_{\bar{p}}}(E_p,E_{\bar{p}})$ and the interstellar CR proton energy spectrum
\begin{equation}
     \label{source}
        q_{\bar{p}} ^{pp}(E_{\bar{p}})= \int^{+\infty}_{E_{\rm th}} \frac{d\sigma_{p \, p \rightarrow \bar{p}}}{dE_{\bar{p}}}(E_p,E_{\bar{p}}) n_{H}(4\pi\Phi_p(E_p)) dE_p,
    \end{equation} 
    
where $n_{H}$ is the ISM hydrogen density, 
$\Phi_p$ is the CR proton flux, $E_p$ and $E_{\bar{p}}$ are the CR proton and antiproton energies, and
$E_{\rm th}$ the production threshold energy equal to $7\,m_p$. 
The overall ISM composition is H:He:C=1:0.1:$5 \times 10^{-4}$ cm$^{-3}$ \cite{1985ApJS...57..173M}.
Whenever needed for illustrative purposes,  we will fix $\Phi_p$ to the fit to the preliminary AMS-02 data \cite{proton_AMS02} reported in~\cite{DiMauro:2014iia}.
\newline
The differential cross section $d\sigma_{pp \rightarrow \bar{p}}/dE_{\bar{p}}$ is in turn the integral over the angle $\theta$ 
between the incoming proton and the final state antiproton momenta
\begin{equation}
     \label{sigma}
       \frac{d\sigma_{pp\to\bar{p}}}{dE_{\bar{p}}}(E_p,E_{\bar{p}}) = 
	2\pi p_{\bar{p}} \int^{\theta_{\rm max}}_{\theta_{\rm min}} E_{\bar{p}}\frac{d^3\sigma}{dp^3_{\bar{p}}} d(-\cos{\theta}),
\end{equation} 
where $p_{\bar{p}}=\sqrt{E^2_{\bar{p}}-m^2_{p}}$, $\theta_{\rm min}=0^{\circ}$ and the expression for $\theta_{\rm max}$ is 
given by Eq.\eqref{eq:thetamaxdef} in Appendix~\ref{sec:Kinematics}. 
The integral in Eq.\eqref{sigma} is computed in the galactic frame at fixed antiproton energy $E_{\bar{p}}$. Its integrand 
represents the ``Lorentz-invariant distribution function'' 
\footnote{For simplicity, in the rest of this work we will interchange this term with ``cross section''.}
for  the process $p + p \rightarrow \bar{p} + X$, i.e. the inclusive antiproton production. 
Inclusive cross sections for processes of the form $a+b \rightarrow c+X$ can be described in terms 
of  the Lorentz-invariant distribution function
\begin{equation}
f(a+b \rightarrow c+X) = E_c \frac{d^3\sigma}{dp_c^3} = \frac{E_c}{\pi} \frac{d^2\sigma}{dp_L dp_T^2} = \frac{d^2\sigma}{\pi dy dp_T^2}\,,
\label{eq:distributiongeneral}
\end{equation}
where $p_L$, $p_T$ and $y$ are respectively the longitudinal and transverse momentum and the rapidity of particle $c$. Traditionally,
the independent variables most frequently used to parameterise this quantity are
\begin{itemize}
\item the centre-of-mass (CM) energy $\sqrt{s} = \sqrt{2 m_p (E_p+m_p)}$, which is uniquely fixed by the total incident proton energy in the lab frame, $E_p$;
\item $p_T$,  the antiproton transverse momentum;
\item the so-called ``radial scaling" variable $x_R$, defined as
\begin{equation}
x_R = \frac{E_{\bar{p}}^*}{E_{\bar{p}, \mathrm{max}}^*} \label{eq:xRdef}\,,
\end{equation}
where $E_{\bar{p}}^*$ is the antiproton energy and $E_{\bar{p}, \mathrm{max}}^*$ is the maximal energy it can acquire, both defined in the CM frame.
The maximal antiproton energy is  (see Appendix~\ref{sec:Kinematics} for details):
\begin{equation}
     \label{sqrtslab}
       E^{*}_{\bar{p},\rm{max}} = \frac{s-8 m^2_p}{2 \sqrt{s}}\,,
\end{equation} 
which---from the condition $  E^{*}_{\bar{p},\rm{max}} \geq m_p$---also implies the threshold energy for the incident proton in the lab frame,
$E_p\geq 7\,m_p$. 
\end{itemize}

The inclusive antiproton production cross section cannot be computed from first principles.
Our primary goal in this work is to provide reliable estimates for the magnitude and the uncertainties of the invariant distribution
\eqref{eq:distributiongeneral}. Our results will be presented mostly in the form
of suitable fitting functions. However, we also want to test how reasonable 
the Ansatz of the chosen functional form(s) is.  To that purpose, we will also compare the fitted functions to an ``agnostic'' 
spline interpolation of the data, which only requires a smooth, piecewise functional dependence. We will mainly focus on
antiprotons with energies ranging from a few GeV and up to ${\cal{O}}(1)$ TeV, with the upper limit of this interval corresponding
roughly to the highest energy that can be probed by AMS-02 and the lower one to the point where astrophysical uncertainties become
so large that they constitute the dominant limiting factor in CR studies, a point which we will also briefly comment upon in 
section \ref{sec:therest}.

\renewcommand{\arraystretch}{1.1}
\begin{table*}
\begin{center} 
\begin{tabular}{|c|c|c|c|} 
\hline \hline 
Experiment & $\sqrt{s}$ (GeV) & $p_T$ (GeV) & $x_R$ \\ 
 \hline 
Dekkers \textit{et al},    CERN 1965 \cite{Dekkers:1965zz}     
& $6.1$, $6.7$                            & $(0., 0.79)$ & $(0.34, 0.65)$ \\
Allaby \textit{et al},     CERN 1970 \cite{Allaby:1970jt}      
& $6.15$                                     & $(0.05, 0.90)$ & $(0.40, 0.94)$ \\
Capiluppi \textit{et al},  CERN 1974 \cite{Capiluppi:1974rt}   
& $23.3$, $30.6$, $44.6$, $53.0$, $62.7$  & $(0.18, 1.29)$ & $(0.06, 0.43)$ \\
Guettler \textit{et al},   CERN 1976 \cite{Guettler:1976ce}    
& $23.0$, $31.0$, $45.0$, $53.0$, $63.0$  & $(0.12, 0.47)$ & $(0.036, 0.092)$ \\
Johnson \textit{et al},    FNAL 1978 \cite{Johnson:1977qx}     
& $13.8$, $19.4$, $27.4$                  & $(0.25, 0.75)$ & $(0.31, 0.55)$ \\
Antreasyan \textit{et al}, FNAL 1979 \citep{Antreasyan:1978cw} 
& $19.4$, $23.8$, $27.4$                  & $(0.77, 6.15)$ & $(0.08, 0.58)$ \\
BRAHMS,                    BNL  2008 \cite{Arsene:2007jd}     
 & $200$                                      & $(0.82, 3.97)$ & $(0.11, 0.39)$ \\
NA49,                      CERN 2010 \cite{Anticic:2009wd}     
& $17.3$                                     & $(0.10, 1.50)$ & $(0.11, 0.44)$ \\
\hline \hline
\end{tabular} \caption{Datasets used in our analysis along with their corresponding $\sqrt{s}$ values and $(p_T, x_R)$ regions.}
\label{tab:datasets}
\end{center} 
\end{table*}
\renewcommand{\arraystretch}{}

\subsection{Experimental data}\label{sec:experiments}

\begin{figure*}[t] 
\begin{center}
\includegraphics[width=1.0\columnwidth]{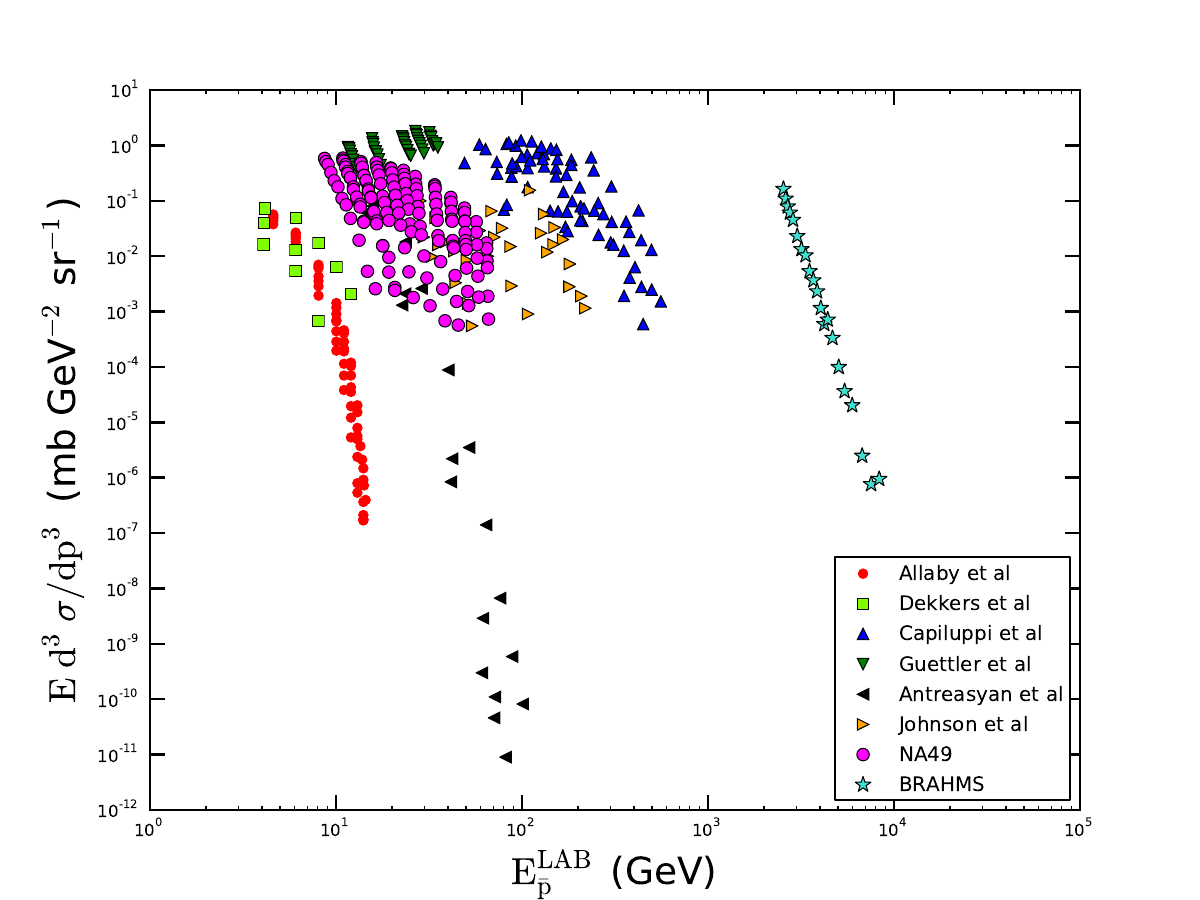} 
\includegraphics[width=1.0\columnwidth]{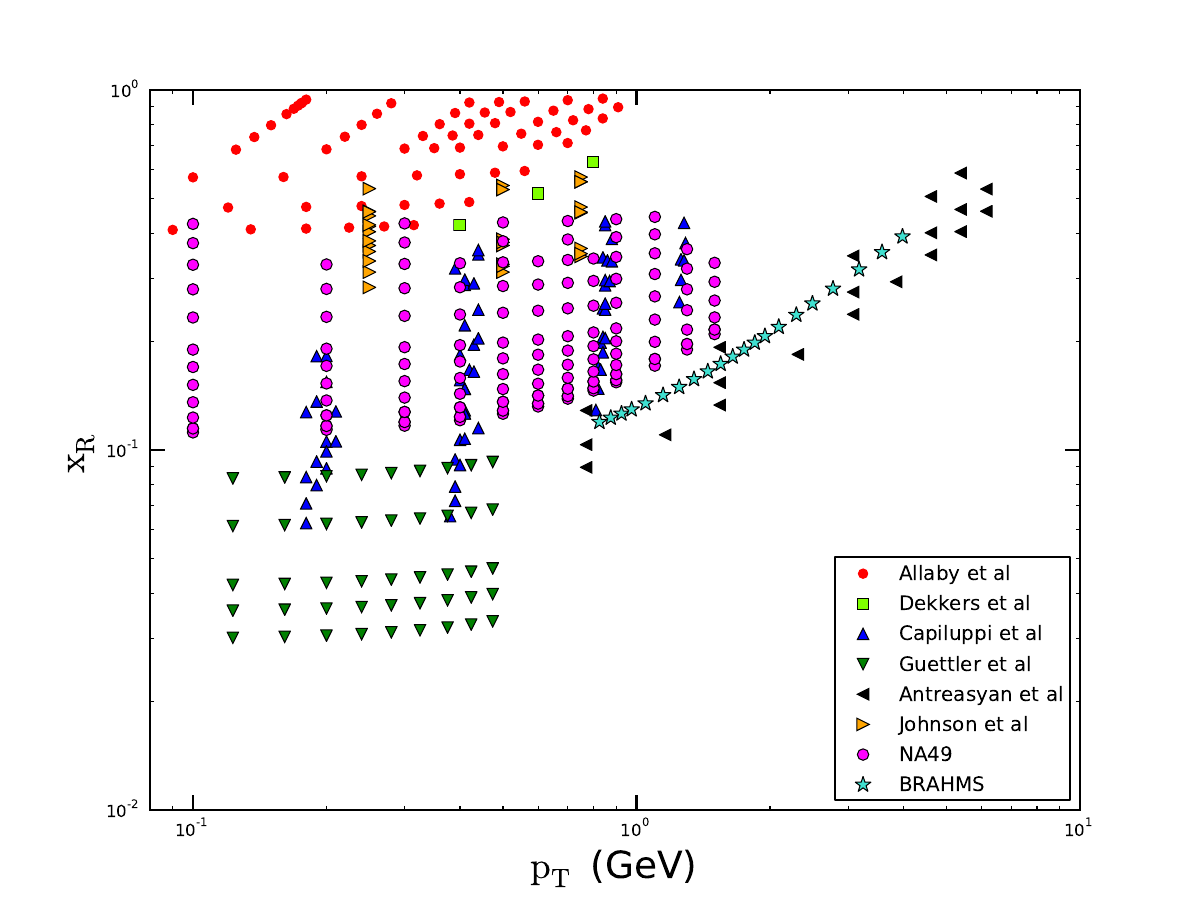} 
\end{center}
\caption{The data on $d^3\sigma_{p \, p \rightarrow \bar{p}}/dp^3$ employed in our analysis are reported as a function of 
$E^{\rm LAB}_{\bar{p}}$ (left panel) and in the $p_T-x_R$ space (right panel). For the data details, see Table \ref{tab:datasets}.}
\label{fig:alldata}
\end{figure*}

In order to estimate the inclusive antiproton production cross section, we consider the 
datasets~\cite{Dekkers:1965zz,Allaby:1970jt,Capiluppi:1974rt,Guettler:1976ce,Johnson:1977qx,Antreasyan:1978cw,Arsene:2007jd,Anticic:2009wd},  
reported in Table \ref{tab:datasets} in terms of the centre-of-mass energy $\sqrt{s}$ and $(p_T, x_R)$ regions (often not rectangular!) 
covered by each experiment. Note that not all experiments provide data in terms of these kinematic variables; 
in those cases, the data were first converted in terms of $\{s,p_T, x_R\}$. We report in Appendix~\ref{sec:Kinematics} the straightforward but 
somewhat lengthy derivation of the transformation equations. 
The data are graphically illustrated in Fig.\ref{fig:alldata}. In the left panel, the cross section is shown as a function 
of $E^{\rm LAB}_{\bar{p}}$, for different combinations of $p_T$ and $x_R$. In the right panel, 
the same data are seen in the $p_T-x_R$ plane. The NA49 data cover wide ranges in both $p_T$ and $x_R$, 
and describe lab antiproton energies from about 8 GeV up to 70 GeV.

Compared to the previous works~\cite{Duperray:2003bd,Tan:1982nc}, the analyses of the NA49~\cite{Anticic:2009wd} and BRAHMS~\cite{Arsene:2007jd} datasets are new to this paper. 
Note that the BRAHMS centre-of-mass energy corresponds to an incident proton energy of roughly $21$ TeV in the lab frame, which lies somewhat beyond the energy region of 
interest for our work.
Given the absence of data for incident proton energies above $\sim 200$ GeV, however, we have included this dataset since it can help in guiding the high-energy
extrapolation of the fit to physical values. It is worth stressing that in the more interesting tens of GeV region for the antiproton laboratory 
energy, the major impact will be provided by far by the NA49 data. 

Another important conceptual issue concerns the possibility to combine data---whose quality and robustness of error assessment is very diverse---in a global fit. 
There is no simple answer to this question: on one hand there are some systematic effects that are certainly present in the old data and hard to estimate and correct for. A known
example is provided by the  {\it feed-down} effect. A significant fraction of antiproton production (easily of ${\cal O}$(20\%)) comes from strange 
hyperon ($\Lambda$, $\Sigma$) decays, whose decay lengths are comparable to or larger than length scales of current micro-vertex detections or precision tracking. 
This effect was taken into account in the NA49 data analysis, where the contribution from hyperons has been subtracted from the
measured yields\footnote{H. G. Fischer, private communication}. For older experiments, no such correction was performed:  
while in some cases---as for the CERN ISR---it may be argued that reasonable estimates make the expected correction negligible, for fixed-target experiments 
covering an extended range of lab momenta the situation is somewhat more complicated. 
No a priori correction has been applied in the following for this effect, especially since ex-novo simulations of trajectories through the detectors and the collimators would be 
needed for robust estimates. For a semi-quantitative discussion, we address the reader to~\cite{Anticic:2009wd}. However, in deriving global fits, we shall allow for
experiment-dependent renormalizations, which may account (at least in an averaged way) for such a correction, see below.

On the other hand, relying only on contemporary data, notably NA49, means having the invariant cross section
at only one point in $\sqrt{s}$, i.e. at one beam energy. Then, in order to obtain the general cross section, one has to supplement the data with some additional
theoretical assumption, such as the {\it scaling} hypothesis~\cite{Taylor:1975tm}, namely that the cross section only depends on $p_T$ and $x_R$. 
While this behaviour is expected to be approximately respected, notably at high $\sqrt{s}$, its quantitative accuracy can only be gauged by comparison with
experimental data. 
For this reason we decided to apply both strategies and to use either fits or interpolations, to all datasets or to NA49 only, 
with or without the scaling hypothesis, to assess the importance of these effects. 

\subsection{Method}\label{sec:method}

Our fits were performed with the {\sf MINUIT} minimization package. 
Let us denote with $k=1,\ldots, L$ the different experimental datasets, with $i_k$ the $i$-th point of the dataset $k$ and 
let ${\bf C}$ be the vector of the cross section parameters.
The fitting procedure consists of varying the values of the cross section parameters ${\bf C}$, comparing the theoretical cross section $F(s_{i_k}, x_{i_k},p_{i_k};{\bf C})$ with the data $f_{i_k}(s_{i_k}, x_{i_k},p_{i_k})$ and finally finding the minimum of ${\chi }^2({\bf C})$ function defined below. This procedure gives the best-fit configuration ${\bf C}_{\rm{best}}$ 
with the corresponding $1\sigma$ errors $\sigma_{{\bf C}}$. 
We define the $\chi^2({\bf C})$ function to be minimised in the fitting procedure in the following way:
\begin{equation}
\chi^2({\bf C})=\chi^2_{\rm stat}({\bf C})+\chi^2_{\rm sys}
\end{equation}
where
\begin{equation}
\chi^2_{\rm sys}=\sum_{k=1}^L\frac{(\omega_k-1)^2}{\epsilon_k^2}\,,
\end{equation}
and
\begin{equation}
\chi^2_{\rm stat}({\bf C})=\sum_{k=1}^L\sum_{i_k}\frac{(\omega_k f_{i_k}-F(s_{i_k}, x_{R_{i_k}},p_{T_{i_k}},{\bf C}))^2}{\omega_k^2\sigma_{i_k}^2}\,.
\end{equation}
In the equations above, $\epsilon_k$ is a systematic overall scale error of the dataset $k$ (either quoted in the experimental paper, or assumed conservatively to be
of the same order of the statistical one if this information is not available, notably for older references); the parameter $\omega_k$ renormalises
the dataset $k$ and is determined consistently by the global fit: of course, large renormalizations with respect to $\epsilon_k$ are disfavoured by the large penalties
to be paid in the global analysis; $\sigma_{i_k}$ is the statistical error on the data point ${i_k}$, 
while the factor $(\omega_k f_{{i_k}}-F)/\omega_k$ is the difference between
experimental values $f$ (accounting for a possible renormalization $\omega_k$, unique for each dataset) and the fitting function $F$, which depends on the independent variables
described above, and on the vector of fitting parameters ${\bf C}$.

This method is the most natural generalization of the {\it unbiased} one presented in~\cite{D'Agostini:1993uj} (see therein equation (3) and section 4) and it has already 
been successfully used in the past for other astroparticle physics analyses involving combinations of different
datasets, as for instance nuclear reaction rates in primordial nucleosynthesis~\cite{Serpico:2004gx}.
\\
\\
Passing on to the data interpolations against which we will be comparing our fitting procedure results, 
one difficulty lies with the fact that $3$-dimensional interpolation of scattered data is a non-trivial problem in contemporary numerical
analysis, with very few (if any) relevant publicly available tools. In order to tackle this issue, in our interpolations we will be making the 
assumption that the invariant distribution \eqref{eq:distributiongeneral} scales with $\sqrt{s}$ only through an overall 
multiplicative dependence on the inelastic $pp$ scattering cross section $\sigma_{\rm in}$. Under this assumption, by dividing 
the experimental data with $\sigma_{\rm in}$ we obtain a $\sqrt{s}$-invariant set of points for which 
only a $2$-dimensional interpolation is needed. Besides, as a by-product of our analysis 
we have re-evaluated the inelastic cross section as described in
Appendix~\ref{sec:sigmainel}.

The interpolations were performed by means of the {\sf Python} routine {\sf SmoothBivariateSpline} contained in the {\sf scipy} library, 
choosing piecewise cubic polynomials as interpolating functions. Note that the routine does not actually perform an exact interpolation, 
rather finds a compromise between the smoothness of the interpolating function and the closeness to the experimental data.

Estimating a statistically meaningful uncertainty band in this approach is fairly tricky. What we did was to consider each experimental best
determination and error as the mean and standard deviation of a gaussian probability distribution of the cross section at that point. 
We then sampled these
distributions accordingly, thus creating a large number of pseudo-experimental points. Each set of points is then interpolated 
(and, depending on the variable one is interested in, eventually integrated over $\cos\theta$ and the proton incident energy/proton flux) to
obtain an ``envelope band'' for the quantity of interest. The average between maximum and minimum of the envelope at each point 
then defines an ``average interpolation curve''.

\section{Results}\label{sec:results}
\subsection{Validation of fitting method}\label{sec:validation}

\begin{figure}[!th] 
\begin{center}
\includegraphics[width=0.99\columnwidth]{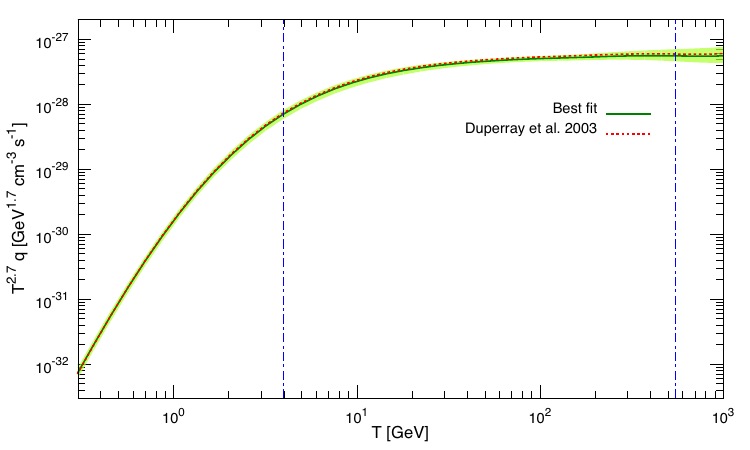} 
\end{center}
\caption{Comparison of the source term for antiproton production in $pp$ collisions as obtained in~\cite{Duperray:2003bd} 
(see however text for a correction in their table) 
and in this work, by re-fitting the same datasets with the same functional form, with our nominal 3$\sigma$ statistical error band. 
Vertical dot-dashed lines show the domain of energy actually covered by the experiments analysed.}
\label{fig:compDup}
\end{figure}

As a preliminary exercise, and in order to validate our kinematical data conversion and fitting routines, we checked if the fit of Eq.~(6)
in~\cite{Duperray:2003bd} is reproduced, of course restricting ourselves to the datasets 
available at the time \cite{Dekkers:1965zz,Allaby:1970jt,Capiluppi:1974rt,Guettler:1976ce,Johnson:1977qx,Antreasyan:1978cw}. 
The parameterization of the invariant cross section is in this case: 
\begin{eqnarray}
E \frac{d^3\sigma}{dp^3} &=& \sigma_{\rm in}(s) (1-x_R)^{D_1} e^{-D_2 x_R} \nonumber \\
&&\left[ D_3 (\sqrt{s})^{D_4} e^{-D_5 p_T} +  D_6 e^{-D_7 p_T^2} \right]\,,
\label{eq:DuperrayParametrization}
\end{eqnarray}
where $\sigma_{\rm in}$ is the total inelastic cross section for $pp$ collisions for which here, and only here, we used the
parameterization adopted in~\cite{Duperray:2003bd}
\begin{equation}
\sigma_{\rm in}(s)  = \sigma_0 \left[1 -0.62\, e^{-\frac{E_{\rm inc}(s)}{200}} \sin{\left(\frac{10.9}{E^{0.28}_{\rm inc}(s)}\right)}\right]\,,
\label{eq:sigmainDuperray}
\end{equation}
where $E_{\rm inc}(s)$ is the incident kinetic energy in MeV defined as $E_{\rm inc}(s) = s/(2m_p)-2 m_p$ and $\sigma_0=44.40$ mbarn.
We show in Tab.~\ref{tab:duparry} our best fit values and 1$\sigma$ errors for the cross section pararemeters $D_i$ in Eq.\eqref{eq:DuperrayParametrization}.
A modest disagreement with the results of~\cite{Duperray:2003bd} was initially found only for $D_1$ and $D_2$, which eventually we could attribute to a typo in the
fitting parameter values reported in their Table V. If we invert $D_1$ with $D_2$, not only do we obtain a very good agreement with our results,
but also a reduced chi-squared $\chi^2_\nu\simeq 3.6$, the same value the authors  quote in their paper. 
By insisting in interpreting literally the values of their table V, we would find $\chi^2_\nu\simeq 9.9$, clearly inconsistent with the value of 3.6.
In Fig.~\ref{fig:compDup} we display the comparison of the best fit and 3$\sigma$ uncertainty band of the source term derived with our best fit values of the 
parameters in Tab.~\ref{tab:duparry}, and the best fit source term with the results reported in~\cite{Duperray:2003bd} with $D_1$ and $D_2$ inverted. 
The two results are essentially indistinguishable.

\begin{table*}[!th]
\begin{center} 
\begin{tabular}{|c|c|c|c|c|c|c|} 
\hline \hline 
$D_1$ (error) & $D_2$ (error) & $D_3$ (error) & $D_4$ (error) & $D_5$ (error) & $D_6$ (error)  & $D_7$ (error)  \\ 
 \hline 
$4.22 (0.66)$  &  $3.435 (0.016) $  & $0.0067 (0.0014)$  &  $0.510 (0.050)$ & $3.609 (0.021)$ & $0.0209 (0.0010)$ & $3.086 (0.083)$ \\
\hline \hline
\end{tabular} \caption{Best-fit values and $1\sigma$ errors for the parameters $D_i$ in Eq.\eqref{eq:DuperrayParametrization} resulting from a fit to the \cite{Duperray:2003bd} dataset.}
\label{tab:duparry}
\end{center} 
\end{table*}

\subsection{Analysis of NA49 data}\label{sec:NA49}

\begin{figure*}[t] 
\begin{center}
\includegraphics[width=0.99\columnwidth]{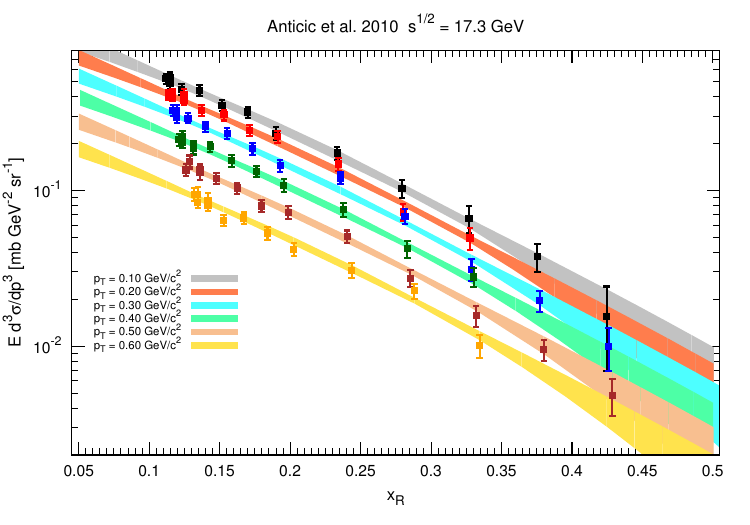} 
\includegraphics[width=0.99\columnwidth]{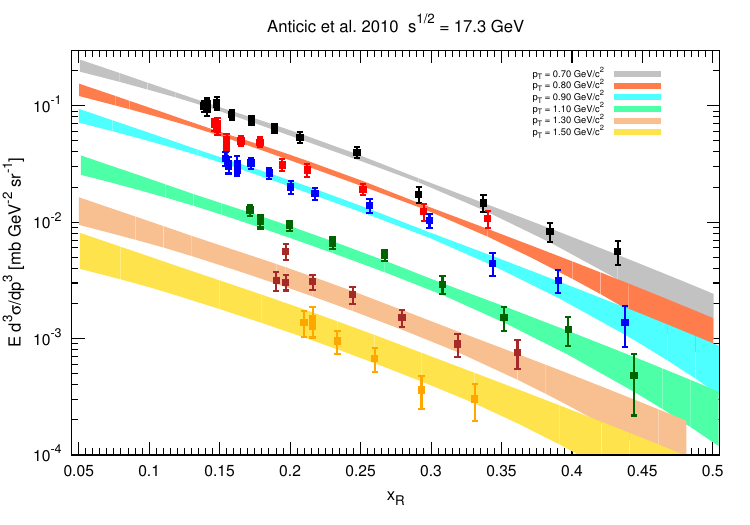} 
\end{center}
\caption{Comparison between NA49 data with the fitting function of Eq.\eqref{eq:NewParametrization1}, see Table~\ref{tab:NA49only}, 
with $3\,\sigma$ error bands. For clarity, the data and the theoretical curves at each $p_{T}$ value have been 
multiplied by a factor of $0.9^{n_{p_T}}$, where $n_{p_T}$ is the  integer counting the  $p_T$, from lower to higher
($i.e.$: for $p_T$= 0.60 GeV/$c^2$ the rescaling is $0.9^6$).}
\label{fig:NA49fitdata}
\end{figure*}

Once our routines validated, we proceed first with fits to the NA49 dataset only. We use the functional form 
\begin{eqnarray}
E \frac{d^3\sigma}{dp^3} = & \sigma_{\rm in}(s) (1-x_R)^{C_1} e^{-C_2 x_R}  \nonumber  \\
& \left[ C_3\, e^{-C_4 p_T} + C_5 \, e^{-C_6 p_T^2} \right]\,,
\label{eq:NewParametrization1}
\end{eqnarray}
where $\sigma_{\rm in}(s)$ is defined in Appendix~\ref{sec:sigmainel}, 
Eq.\eqref{eq:sigmappinelparam}. This functional form is a simplified version of the standard parametrization proposed in \cite{KMNref}
(it has four parameters less), which we found to provide
an accurate and more compact description of the data. Note that we implicitly assume some form of scaling, in that the only dependence
on $s$ is given by the overall multiplication with the inelastic cross section.
The best-fit values and the $1\sigma$ errors are reported
in Table~\ref{tab:NA49only}, with the corresponding fit having a reduced chi-square $\chi^2_\nu=1.3$ for $137$ degrees of freedom.
\begin{table*}[!th]
\begin{center} 
\begin{tabular}{|c|c|c|c|c|c|} 
\hline \hline 
$C_1$ (error) & $C_2$ (error) & $C_3$ (error) & $C_4$ (error) & $C_5$ (error) & $C_6$ (error)  \\ 
 \hline 
$7.56 (1.15)$  &  $0.245 (0.148) $  & $0.0164 (0.0025)$  &  $2.37 (0.13)$ & $0.0352 (0.0020)$ & $2.902 (0.059)$ \\
\hline \hline
\end{tabular} \caption{Best-fit parameters and $1\sigma$ errors to the NA49 data~\cite{Anticic:2009wd} with Eq.\eqref{eq:NewParametrization1} .}
\label{tab:NA49only}
\end{center} 
\end{table*}
The comparison between data and fitted function (along with the corresponding 3$\sigma$ bands) is presented in Fig.~\ref{fig:NA49fitdata}. 
We see that the data are well represented by the fitting function, Eq.\eqref{eq:NewParametrization1}, for all the $p_T$ and $x_R$ values.

\begin{figure}[!th] 
\begin{center}
\includegraphics[width=1.0\columnwidth]{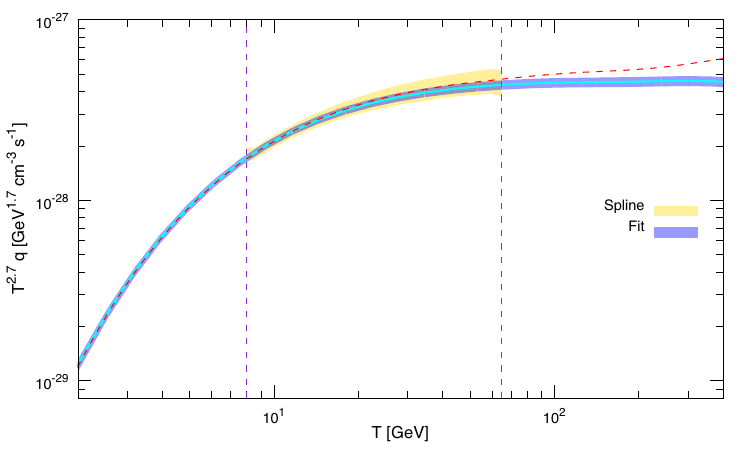} 
\end{center}
\caption{Comparison between the function Eq.\eqref{eq:NewParametrization1} fitted to NA49 data, with 3$\sigma$ error band (solid curve with cyan/blue shaded band), 
and interpolated curve (dashed red), with the interpolation envelope band, yellow/light shading.
The dashed vertical lines correspond to the equivalent antiproton energy sampled by the NA49 experiment, where an interpolation is actually justified.}
\label{fig:figNA49int}
\end{figure}

Next, we checked that the chosen fitting formula does not impose too strong a theoretical bias. To that purpose, as described in 
paragraph \ref{sec:method}, we performed an ``educated'' interpolation of the data by dividing the
datapoints by $\sigma_{\rm in}(s_{\rm NA49})$ and assuming that the resulting function is independent of $s$. 
The final function which is obtained by re-multiplying by $\sigma_{\rm in}(s)$ thus still has a dependence on $s$, 
albeit a trivial one, via the overall factor $\sigma_{\rm in}(s)$. 
The comparison between our fitting and interpolating procedures is  shown in yellow in~Fig.~\ref{fig:figNA49int}.
The vertical lines correspond to the equivalent antiproton energy spanned by the NA49 experiment, where an interpolation is meaningful. 
In order to obtain a reasonable interpolation outside this interval, we supplemented the datasets with ``fake'' points at the boundaries 
of the interpolation grid, with very large errors not to artificially influence the curve, yet sufficient to prevent the numerical 
routine from being driven to extreme functional form interpolations (for example, negative cross sections).
No reasonable error can be however assigned outside the region covered by the data, apart for a lower limit that should be at 
least as large as the maximum relative width of the yellow band. The fact that the average interpolation curve is always 
within $\sim 3\,\sigma$ of the best-fit previously obtained suggests that this $3\,\sigma$ band is roughly representing the 
maximum uncertainty (at least where data exist), accounting not only for statistical errors, but also for possible theoretical biases, acting as additional systematics, related to the choice of the fitting function.

\subsection{Global analysis}\label{sec:global}

Finally, we proceed to the global analysis of all available data on $pp \rightarrow \bar{p} + X$ listed in Table \ref{tab:datasets}.
In this case, given that we wish to describe data referring to quite different $\sqrt{s}$ values and covering different $(p_T, x_R)$ 
regions, it is expected that we will have to introduce some complication with respect to the previous paragraph. In this spirit, we 
tried numerous different functional forms, essentially variations of the standard parametrization proposed in
~\cite{KMNref}. We present here results on our two most successful attempts, which also provide interesting insights on the extrapolation
to regions where data are either scarce or altogether unavailable, a point that we shall discuss in more detail in section 
\ref{sec:discussion}.

As a first step, we used an improved version of Eq.\eqref{eq:NewParametrization1} introducing an explicit dependence on $s$, namely
\begin{eqnarray}
\label{eq:NewParametrization3}
E \frac{d^3\sigma}{dp^3} &=& \sigma_{\rm in}(s) (1-x_R)^{C_1} e^{-C_2 x_R}  \\ 
&&\left[C_3 (\sqrt{s})^{C_4} e^{-C_5 p_T} + C_6 (\sqrt{s})^{C_7} e^{-C_8 p_T^2} \right]. \nonumber
\end{eqnarray}
This parametrization of the cross section is similar to the one proposed in~\cite{KMNref} except for the absence of a  $\sqrt{s}$ 
exponent in the $(1-x_R)^{C_1}$ term. The fit gives a reduced chi-square of $\chi^2_\nu=4.16$, with a number of degrees of freedom of 385. 
The best-fit values and uncertainties are reported in Table \ref{tab:all3}.
We have also checked that considering the exact form as in~\cite{KMNref} we obtain an even worse fit to the data 
($\chi^2_{\nu}=5.6$ with 385 degrees of freedom).
\begin{table*}[!th]
\begin{center} 
\begin{tabular}{|c|c|c|c|c|c|c|c|} 
\hline \hline 
$C_1$ (error) & $C_2$ (error) & $C_3$ (error) & $C_4$ (error) & $C_5$ (error) & $C_6$ (error) & $C_7$ (error) & $C_8$ (error) \\ 
 \hline 
4.499(0.040) & 3.41(0.11) & 0.00942(0.00083) & 0.445(0.027) & 3.502(0.018) & 0.0622(0.0086) & -0.247(0.049) & 2.576(0.027) \\
\hline \hline
\end{tabular} 
\caption{Best-fit values and $1\sigma$ errors for the parameters $C_i$ in Eq.\eqref{eq:NewParametrization3} derived with a fit to all datasets.}
\label{tab:all3}
\end{center} 
\end{table*}

Motivated by the relatively poor quality of the fit, we tried an extended version of Eq.\eqref{eq:NewParametrization3}, namely
\begin{eqnarray}
E&& \frac{d^3\sigma}{dp^3} = \sigma_{\rm in}(s) (1-x_R)^{C_1} e^{-C_2 x_R}  \left | [ C_3 (\sqrt{s})^{C_4} e^{-C_5 p_T} + \right.  \nonumber \\ 
 &&\left.   C_6 (\sqrt{s})^{C_7} e^{-C_8 p_T^2} + C_9 (\sqrt{s})^{C_{10}} e^{-C_{11} p_T^3}] \right | ,
\label{eq:NewParametrization2}
\end{eqnarray}
where the absolute value simply prevents the function from becoming negative in some corners of parameter space. Compared to
the previous function, this one
further contains the exponential of a cubic function of $p_T$ and an additional $s$-dependence.
The best-fit parameters for 
Eq.\eqref{eq:NewParametrization2} are reported in Table~\ref{tab:all}. This parametrization yields a somewhat better $\chi^2_\nu=3.30$ 
for 382 degrees of freedom.
In order to test the validity of the scaling hypothesis, 
we obtain that the fit to the same data with Eq.\ref {eq:NewParametrization2}  containing no dependence on s but not in $\sigma_{\rm in}(s)$, 
gives a reduced $\chi^2_\nu=4.82$.
\begin{table*}[!th]
\begin{center} 
\begin{tabular}{|c|c|c|c|c|c|} 
\hline \hline 
$C_1$ (error) & $C_2$ (error) & $C_3$ (error) & $C_4$ (error) & $C_5$ (error) & $C_6$ (error) \\ 
 \hline 
4.448(0.035) & 3.735(0.094) & 0.00502(0.00036) & 0.708(0.019) & 3.527(0.014) & 0.236(0.024) \\
\hline \hline
\end{tabular} 
\begin{tabular}{|c|c|c|c|c|c|} 
\hline \hline 
$C_7$ (error) & $C_8$ (error)& $C_9$ (error) & $C_{10}$ (error) & $C_{11}$ (error) \\ 
 \hline 
 -0.729(0.036) & 2.517(0.027) & $-1.822(0.009) 10^{-11}$ & 3.527(0.022) & 0.384(0.021) \\
\hline \hline
\end{tabular} 
\caption{Best-fit values and $1\sigma$ errors for the parameters $C_i$ in Eq.\eqref{eq:NewParametrization2} derived with a fit to all datasets.}
\label{tab:all}
\end{center} 
\end{table*}

\begin{figure*}[!th] 
\begin{center}
\includegraphics[width=0.9\columnwidth]{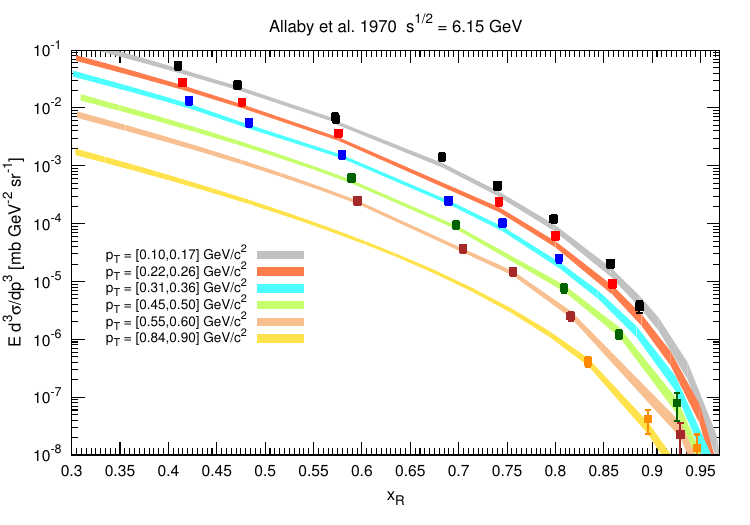} 
\includegraphics[width=0.9\columnwidth]{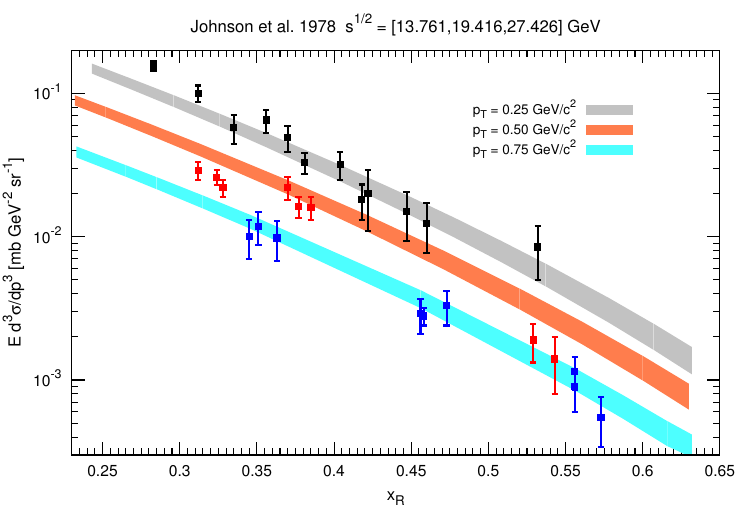} 
\includegraphics[width=0.9\columnwidth]{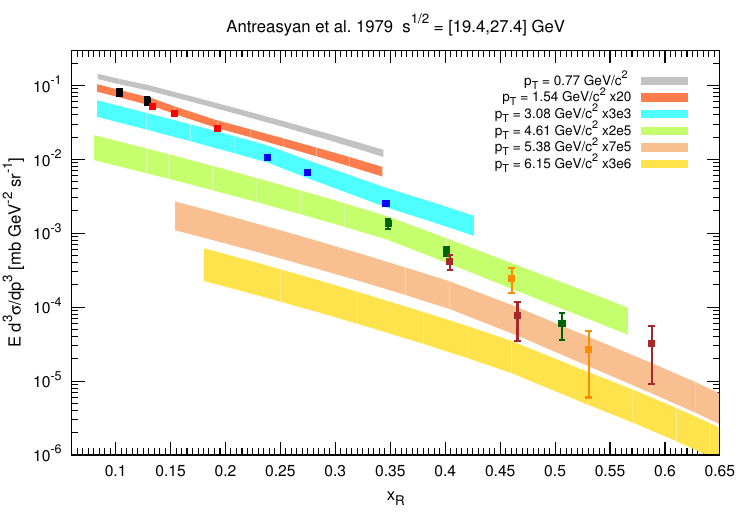} 
\includegraphics[width=0.9\columnwidth]{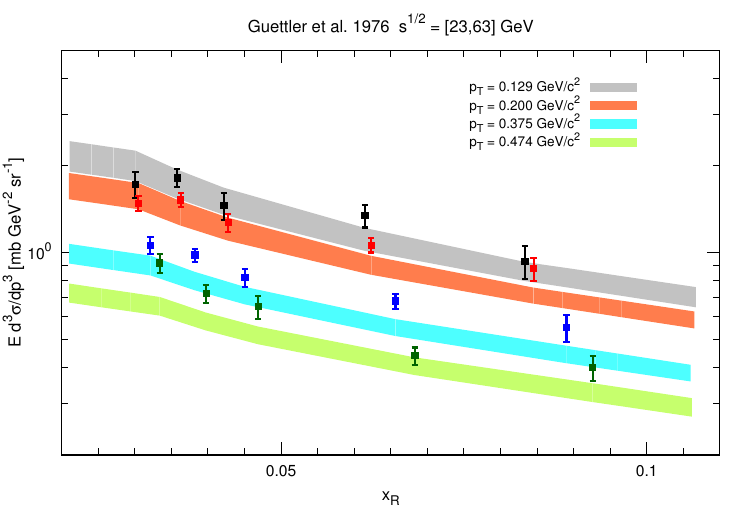} 
\includegraphics[width=0.9\columnwidth]{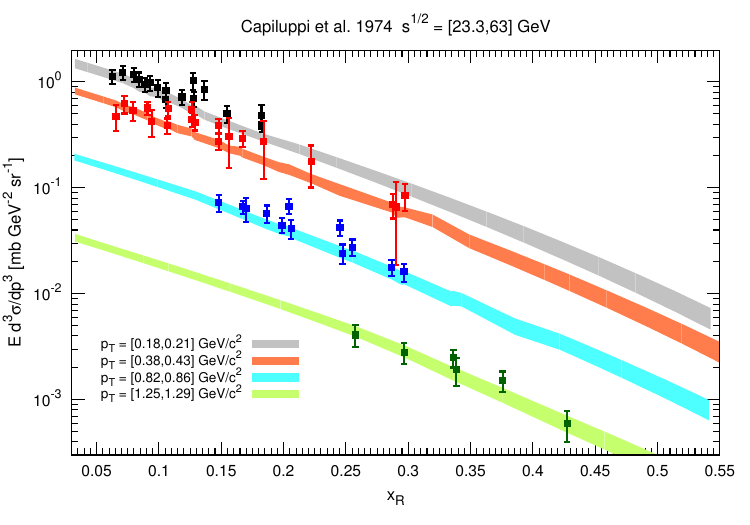} 
\includegraphics[width=0.9\columnwidth]{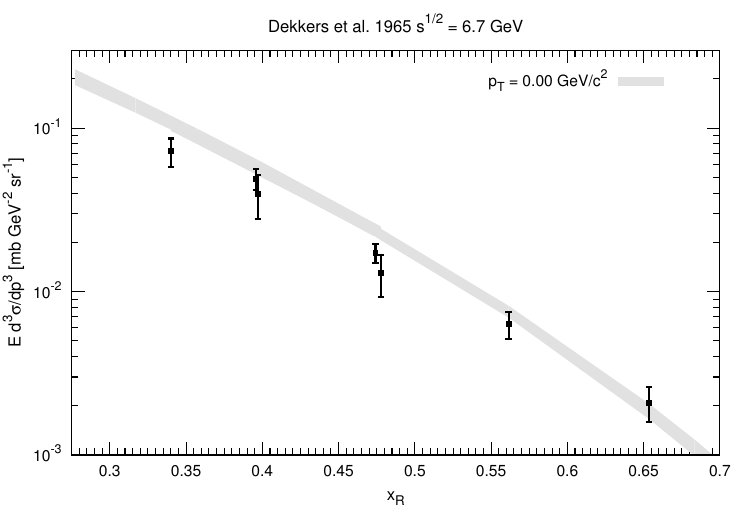} 
\includegraphics[width=0.9\columnwidth]{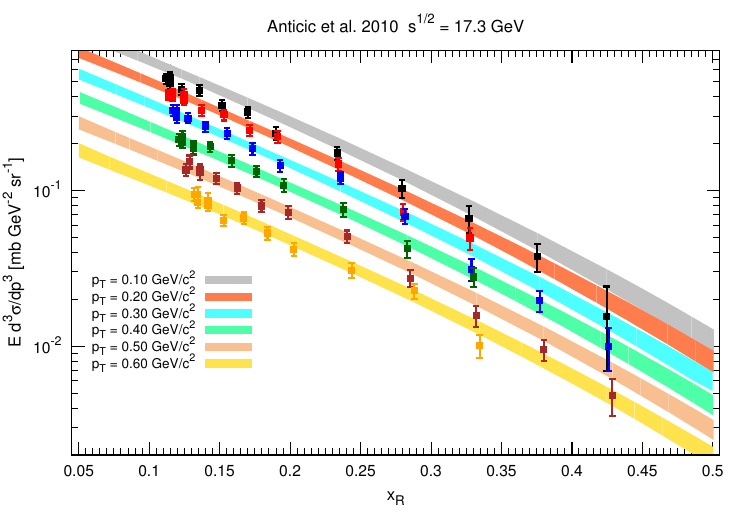} 
\includegraphics[width=0.9\columnwidth]{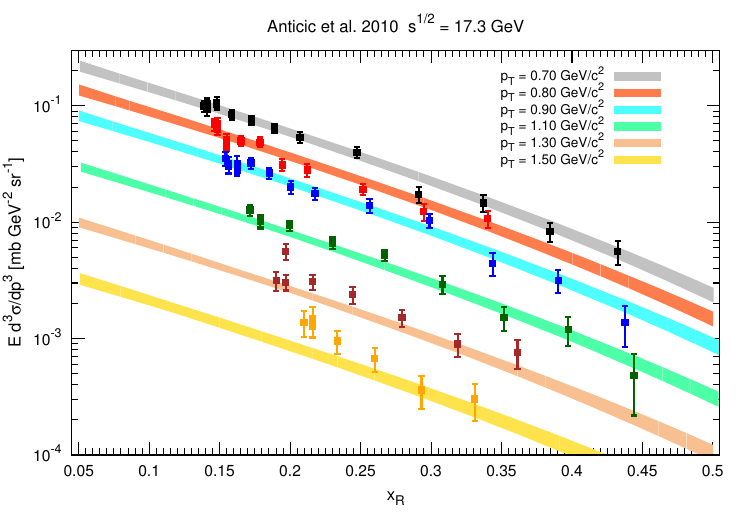} 
\end{center}
\vspace{-2mm}
\caption{Differental cross section for antiproton production in $pp$ scattering, as a function of $x_R$, for different $p_T$ values. 
The curves refer to the  3$\sigma$ uncertainty band around the best fit obtained with a fit of Eq.\eqref{eq:NewParametrization2} 
to the datasets in Tab.~\ref{tab:datasets}. The data, from top left to bottom right, are from 
\cite{Allaby:1970jt,Johnson:1977qx,Antreasyan:1978cw,Guettler:1976ce,Capiluppi:1974rt,Dekkers:1965zz,Anticic:2009wd}. 
For the sake of clarity, the data from \cite{Allaby:1970jt} and \cite{Anticic:2009wd} and the relevant theoretical curves at each $p_{T}$ value have been 
rescaled by a factor $0.6^{n_{p_T}}$ and $0.9^{n_{p_T}}$, respectively, as described in Fig.\ref{fig:NA49fitdata}.}
\label{fig:compgeneral}
\end{figure*}

The improved value of $\chi^2_\nu$ is obtained at the expense of some rescaling of the datasets. With respect to our best result
given by Eq.\eqref{eq:NewParametrization2},  the measurements reported in 
\cite{Allaby:1970jt,Johnson:1977qx,Antreasyan:1978cw,Guettler:1976ce,Capiluppi:1974rt,Dekkers:1965zz,Anticic:2009wd,Arsene:2007jd}
are renormalised respectively by factors $\omega_k$ of $\{0.87,1.04,1.16,0.98,0.95,1.13,1.02,1.16\}$.
Therefore the NA49 data \cite{Anticic:2009wd}, which represent the bulk of the fitting procedure, are renormalised  
by a negligible $\sim 2\%$ while \cite{Allaby:1970jt,Antreasyan:1978cw,Dekkers:1965zz,Arsene:2007jd} by more than $10\%$. Interestingly, the largest renormalization 
value is $16\%$ for the BRAHMS dataset \cite{Arsene:2007jd} giving a factor of 1.16, still not particularly significant given the statistical errors,
yet perhaps indicative of some ``theoretical error'' effects which become more prominent when an agreement over a  large energy range is demanded. 
We display in Fig.~\ref{fig:compgeneral} the comparison of the cross section best fit and 3$\sigma$ uncertainty band 
according to Eq.\eqref{eq:NewParametrization2} with the datasets \cite{Allaby:1970jt,Johnson:1977qx,Antreasyan:1978cw,Guettler:1976ce,Capiluppi:1974rt,Dekkers:1965zz,Anticic:2009wd}. 
We omit the comparison with the BRAHMS results, because in this case the cross section has only been measured 
along a line in the $(p_T,x_R)$ space (see Fig.~\ref{fig:alldata}). 
Nevertheless, the difference between our best fit cross section and the data 
in \citep{Arsene:2007jd} is at most $\sim 30\%$.
We see that most of the data are well reproduced by the fitting function of Eq.\eqref{eq:NewParametrization2} at all $p_T$ values. 
This is true in particular for the NA49 data, except for a slight overestimation at the lowest $p_T$ value. We have however checked that a 
$20\%$ shift in the  differential cross section for $p_T<0.15$ has a negligible effect on the antiproton source term (less than $5\%$). 

We then repeated the interpolation analysis, previously only performed for NA49, for the entire dataset. In this case, the parameter space
coverage is such that there is no need to supplement the dataset with ``fake'' points, as previously done for the NA49 data alone. 
The spline method results in this case are, thus, fully data-driven, modulo our implicit assumption concerning the cross section $\sqrt{s}$ 
scaling according to an overall factor $\sigma_{\rm in}(s)$.

\begin{figure}[!th] 
\begin{center}
\includegraphics[width=1.0\columnwidth]{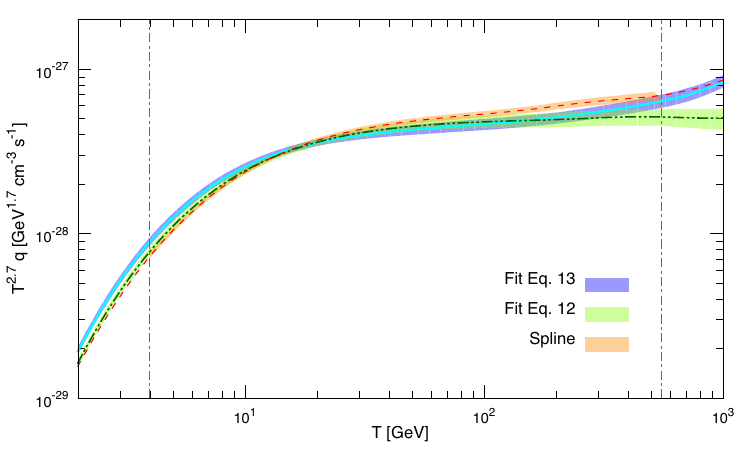} 
\end{center}
\caption{Comparison between fitted function of Eq.\eqref{eq:NewParametrization2} with $3\,\sigma$ band (solid curve with cyan/blue shaded band), 
of Eq.\eqref{eq:NewParametrization3} with $3\,\sigma$ band (dot-dashed curve with green/light green shaded band)
and interpolated curve (dashed red), with  the interpolation envelope band, red/orange shading.
The dashed vertical lines correspond to the equivalent antiproton energy sampled by the global dataset, where an interpolation is in principle meaningful. }
\label{fig:figall}
\end{figure}

In Fig.~\ref{fig:figall} we compare the results obtained for the antiproton source term
through our fits according to the equations \eqref{eq:NewParametrization2} and
\eqref{eq:NewParametrization3}, as well as the estimate based on our spline interpolation. 
The energy range where $pp$ data (except for BRAHMS) are available is bracketed by the vertical lines.
We see that above 10 GeV, 
and within the region where experimental data are available, all three methods yield compatible results. 
At lower and higher energies, however, there is a significant departure of the three estimates. We will discuss the implications
of these results in much more detail in section \ref{sec:discussion}.

\begin{figure}[!th] 
\begin{center}
\includegraphics[width=1.0\columnwidth]{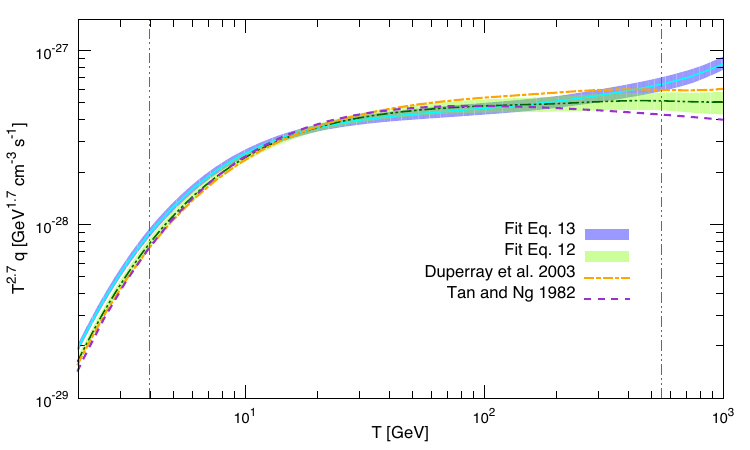} 
\end{center}
\caption{The best fit and 3$\sigma$ uncertainty band source term derived with the fit of Eq.\eqref{eq:NewParametrization3} and Eq.\eqref{eq:NewParametrization2} 
to all datasets in Tab.~\ref{tab:datasets} is shown together with the source term obtained using ~\cite{Duperray:2003bd,Tan:1982nc} cross section parametrizations.}
\label{fig:figcomp}
\end{figure}

In order to compare the results derived in this section to previous published proton-proton cross section 
estimates, we show in Fig.~\ref{fig:figcomp} 
the best fit and 3$\sigma$ uncertainty band source term calculated with our results and the best fit source term derived with the 
parametrizations adopted in ~\cite{Duperray:2003bd,Tan:1982nc}.
In the range of antiproton kinetic energy where data exist, our 3$\sigma$ band is 
marginally compatible
with the parameterization in ~\cite{Duperray:2003bd,Tan:1982nc}, which is overestimated (underestimated) below (above) about 20 GeV.

\section{Contributions from neutrons and nuclei}\label{sec:therest}

In order to obtain the total antiproton source term, two more effects should be taken into account: the effects of nuclear projectiles and targets in the collisions,
 and the yield coming from antineutron production. An exhaustive treatment of both subjects goes beyond our current purposes.
For completeness, however, in the following we summarise the re-scalings of the yield from the $pp\to {\bar p}$  process that are usually adopted in the 
literature to account for both the processes, and some of the issues involved.

Concerning nuclear enhancements (effects of proton-nucleus, nucleus-proton, and nucleus-nucleus collisions), unfortunately very little data
are present, notably none for the most important channels which are the ones involving helium.  
One possible strategy to deduce cross sections for reactions involving helium is to constrain
those of nuclear species for which some data are available, and extrapolate from heavier species to lighter ones, see e.g.~\cite{Duperray:2003bd}.
Given that helium is quite light, however, it has often been considered reliable to deduce the relevant cross sections from rescaling the $pp$ ones either with
semi-empirical formulae or via hadronic models, see e.g.~\cite{Donato:2001ms}.
The most recent dedicated studies were performed on the basis of the  Monte Carlo (MC) model DTUNUC in~\cite{1998ApJ...499..250S} and in~\cite{Donato:2001ms}.
The models implemented in the software are based on the Dual Parton Model~\cite{Capella:1992yb} and the Gribov-Glauber approach for a unified
treatment of soft and hard scattering processes. The former are parameterised according to Regge phenomenology whereas the latter
rely on lowest order perturbative QCD. Eventually, DTUNUC formed the basis of/merged into DPMJET-III (see~\cite{Roesler:2000he} and refs. therein
for further details)~\footnote{Recently, some new theoretical evaluation appeared in the Appendix of~\cite{Kachelriess:2012fz}.}.
\\
A fit of the nuclei enhancement yield of antiprotons found in~\cite{1998ApJ...499..250S} compared to the one in $pp$ collision, is given in~\cite{Moskalenko:2001ya}:
\begin{equation}
Q_{\rm tot}/Q_{\rm pp}=0.12(T_{\bar p}/{\rm GeV})^{-1.67}+1.78\,,
\end{equation}
with the above expression assuming 10\% density ratio of He to H nuclei. Note that this ignores different spectral indices and species-dependent
spectral breaks, which have been reported by some experiments but have not been confirmed by preliminary results of AMS-02. To gauge their possible effect at high-energies,
we address the reader to the brief discussion in~\cite{Donato:2010vm}.

Lacking empirical information for the most relevant channels involving helium, it is hard to assess the accuracy of the previous models. The overall uncertainty 
(on the total source term yield $Q_{\rm tot}$)  was estimated in~\cite{1998ApJ...499..250S} to reach 40\%, from the dispersion of predictions based on different prescriptions, but this 
conclusion is overly pessimistic, since not all the models/evaluations have the same reliability (some were based on obsolete prescriptions, for instance). 
In~\cite{Donato:2001ms} the error estimate was closer to $20-25\%$, provided that the $pp$ cross section does not depart from the Tan \& Ng parameterization~\cite{Tan:1982nc} 
by more than 10\%, which seems to be only marginally compatible with our results.

Since all these prescriptions do not include subtle nuclear effects, it is also likely that the uncertainty at low energies (where they are expected to play a larger role) is significantly larger than at 
high energy. If, in addition, one considers the more complicated astrophysical propagation effects at low-energies 
(reacceleation, convection, solar modulation)~\footnote{It has been clearly shown as different propagation setups can be responsible for a $\sim$50\% min-max dispersion 
in the flux expected at low-energy \cite{2012PhRvD..85l3511E,Cirelli:2014lwa}.}
and the need to correctly 
account for (catastrophic and non-catastrophic) energy  losses, themselves affected by errors, it is clear that below a few GeV's the lower the energy the less 
reliable is the theoretical prediction. 
Most likely, this energy window cannot be used (but very crudely) for astroparticle physics constraints. 

Another correction which is needed to infer the total antiproton flux from $\sigma_{pp\to {\bar p}}$ consists in accounting for the antiproton flux coming from antineutron production.
Traditionally, it has been assumed that  $\sigma_{pp\to {\bar n}}=\sigma_{pp\to {\bar p}}$, so that the previous results have been simply multiplied by a factor 2. 
However, the NA49 collaboration itself~\cite{Fischer:2003xh} has reported an isospin-dependence in studies of secondary yields in $np$ and $pp$
collisions: in $pp$ reactions, there is a significant preference of the positively charged $p{\bar n}$ combination over ${\bar p}n$ (the opposite being true for neutron projectiles).
This results in $\sigma_{pp\to {\bar n}}=\kappa \sigma_{pp\to {\bar p}}$ with $\kappa\simeq 1.5$ around $x_F\sim 0$ (see also Fig. 3 in~\cite{Fischer:2003xh}; $x_F$ is defined in Appendix \ref{sec:Kinematics}), although
the effect depends on $x_F$ to some extent. Given the still rudimentary knowledge of these effects, an energy independent
rescaling $\kappa\simeq 1.3\pm 0.2$ should encompass the data and be a better approximation than the usually assumed $\kappa=1$.
It is clear that addressing these issues is of paramount importance for further reducing the uncertainties in the antiproton source term.


\section{Discussion and conclusions}\label{sec:discussionconclusions}

\subsection{Discussion}\label{sec:discussion}

We now discuss our findings focusing on the global analysis outlined in Sect. \ref{sec:global}. As we can see in Fig.~\ref{fig:figcomp}, for 
antiproton energies lying roughly within the interval $(10, 300)$ GeV, we find that our results on the antiproton source term from proton-proton 
scattering are consistent with previous estimates.
They are moreover stable with respect to reasonable changes in the parametrization choice and in agreement with data-driven 
methods.
These findings can be understood considering that the majority of the data lie in the $T \in (10, 300)$ GeV range, 
where the most reliable estimates of the distribution in Eq.\eqref{eq:distributiongeneral} can be obtained and which, 
even prior to the NA49 and BRAHMS measurements, were already discreetly populated with data.
In this sense, given that the NA49 data are not in contradiction with previous
experimental results, it is expected (and verified) that the estimates presented in Tan \& Ng ~\cite{Tan:1982nc} 
and Duperray \textit{et al} \cite{Duperray:2003bd} are in good agreement with our findings for this energy range. 
Moreover, as long as a reasonable functional form is adopted for the invariant distribution, 
it is more or less bound to predict a comparable source term within this energy range.
The small discrepancies of our spline interpolation and fitting approaches could be likely attributed to the fact that
the interpolation essentially neglects the {\it scaling violation}, while the fits do allow for some flexibility (extra dependence on $s$) 
to accommodate it.

On the other hand, at low and high energies, the relatively small amount of available data 
essentially implies extrapolations of the fits performed principally for $T$ between $8$ and
$300$ GeV. Consequently, moderately different assumptions can yield significantly different results. 
This is demonstrated by the fact that adopting two slightly different parameterizations
while using the same dataset changes the high-energy source term prediction quite dramatically. 
Moreover, these findings are insensitive to
the inclusion or not of the BRAHMS data in the analysis, which means that the results in \cite{Arsene:2007jd} are not 
 sufficient to constrain the high-energy behavior of the invariant distribution and, hence, the antiproton source function. This is due to the fact that the data of~\cite{Arsene:2007jd},  only cover the exponentially suppressed high-$p_T$ region (similarly to the ones of~\cite{Antreasyan:1978cw}), see Fig.~\ref{fig:alldata}. In this sense, 
both the low-energy and high-energy behavior of the invariant distribution remain highly uncertain. Given
that both the spline method and the fit with Eq.\eqref{eq:NewParametrization2} demonstrate a similar trend at high energies, we
believe that making any conclusive statement concerning the high-energy behaviour of the antiproton inclusive cross section
would be risky. This is all the more the case since spline interpolations can be notoriously misleading when extrapolated 
outside data-covered regions.

Whereas in the low-energy regime this point is not very important, given that in any case the secondary antiproton 
flux is dominated by huge uncertainties coming from astrophysical
sources (solar modulation, propagation parameters, antiproton scattering cross sections), it is plausible that in the
region of several hundreds of GeV and higher the main uncertainty is still  due to the antiproton production 
cross section.

\renewcommand{\arraystretch}{1.5}
\begin{table*}[!th]
\begin{center} 
\begin{tabular}{|c|c|c|c|c|c|c|} 
\hline \hline 
 $T$ (GeV) & Eq.\eqref{eq:NewParametrization1} (\% error) & Eq.\eqref{eq:NewParametrization3} (\% error) & Eq.\eqref{eq:NewParametrization2} (\% error) & spline (\% error) & Tan \& Ng & Duperray \textit{et al}  \\ 
 \hline 
$5$     &  $1.23\cdot 10^{-30} (4.9) $ &  $1.47\cdot 10^{-30} (6.1) $  & $1.67\cdot 10^{-30} (5.4)$  &  $1.38\cdot 10^{-30} (2.7)$ & $1.42\cdot 10^{-30} $ & $1.40\cdot 10^{-30}$ \\
$10$    &  $4.31\cdot 10^{-31} (4.2) $ &  $4.87\cdot 10^{-31} (3.0) $  & $5.17\cdot 10^{-31} (4.8)$  &  $4.34\cdot 10^{-31} (2.5)$ & $4.96\cdot 10^{-31} $ & $4.74\cdot 10^{-31} $ \\
$100$   &  $1.70\cdot 10^{-33} (5.9) $ &  $1.82\cdot 10^{-33} (8.7) $  & $1.77\cdot 10^{-33} (6.8)$  &  $2.03\cdot 10^{-33} (3.2)$ & $1.82\cdot 10^{-33} $ & $2.04\cdot 10^{-33}$ \\
$500$   &  $2.42\cdot 10^{-35} (6.2) $ &  $2.82\cdot 10^{-35} (9.5) $  & $3.39\cdot 10^{-35} (8.8)$  &  $3.26\cdot 10^{-35} (5.2)$ & $2.38\cdot 10^{-35}$ & $3.27\cdot 10^{-35}$ \\
$1000$  &  $3.13\cdot 10^{-36} (6.9) $ &  $4.16\cdot 10^{-36} (11) $  & $6.83\cdot 10^{-36} (10)$  &  $7.02\cdot 10^{-36} (5.8)$ & $3.29\cdot 10^{-36}$ & $4.93\cdot 10^{-36}$ \\
\hline \hline
\end{tabular} \caption{Best-fit values and corresponding percentage relative errors for the $pp$-induced source term (in
GeV$^{-1}$cm$^{-3}$s$^{-1}$), for some representative antiproton energies and  different approaches in the data analysis.}
\label{tab:benchmarks}
\end{center} 
\end{table*}
\renewcommand{\arraystretch}{}

We summarise in Table \ref{tab:benchmarks} the $pp$-induced source term
along with the associated percentage uncertainties resulting from our analysis of the NA49 data according to Eq.\eqref{eq:NewParametrization1}, 
our global analysis according to Eqs.\eqref{eq:NewParametrization3} and
\eqref{eq:NewParametrization2}, our spline interpolation method of the full dataset, and the previous estimates 
in \cite{Tan:1982nc} and \cite{Duperray:2003bd},  for a few representative values of the antiproton energy. 
This table simply illustrates the results reported in Figs.
\ref{fig:figall} and \ref{fig:figcomp}: with increasing energy, the different approaches turn from marginally compatible (at the lowest 
energies, few GeV) to fully compatible until,
towards the end of the region for which experimental data are available, they yield very different results. 

Concerning the error estimates, we also point out that the nominal $1\,\sigma$ error band obtained from the $\chi^2$ minimization procedure
is underestimated, for several reasons. In some case where $\chi^2$/dof is close to $1$, as in the fit to NA49 data only with a simple fitting
formula, we showed how the agreement with an interpolation method is only meaningful if roughly a $3\,\sigma$ band is used as typical estimate
of the error. This is the choice we presented in our plots. A similar prescription was found to be more indicative of the real uncertainty, once global fits were performed. 
In this case, the inadequacy of the nominal $1\,\sigma$ error band was already hinted to by the relatively large reduced  $\chi^2$, never smaller than $\chi^2_\nu = 3.30$.
We attribute these results to a combination of factors: i) underestimated experimental errors,  notably in (some of) the older datasets, due to effects
that were neglected as the feed-down we mentioned. ii) inadequacy of any simple functional form tested to describe faithfully the data, especially on a
large dynamic range; iii) some sort of more or less implicit analytical extrapolation assumption in order to achieve coverage of the 
 $3$-dimensional space $(\sqrt{s}, p_T, x_R)$ starting from a discrete set of points. Note that this also applies to interpolation techniques, which 
 for instance rely on some theoretical assumptions such as scaling.
The situation may be certainly improved if high-quality measurements such as the ones provided by NA49 could be extended to a broader dynamic range.

We also stress that outside the regions where data are available, there is no compelling reason for either one of our
results according to equations \eqref{eq:NewParametrization3} and \eqref{eq:NewParametrization2} to be more realistic 
than the other. Whereas the agreement of all of our computations
at intermediate energies hints that the error estimates there is fairly reliable, this is not at
all the case at very low and high energies. 
A more conservative approach is to assume that in this case the error is dominated by the extrapolation uncertainty, 
for which a proxy is given by the region spanned by the ensemble of our approaches, amounting to about $50\%$ at $1$ TeV.

\begin{figure}
\begin{center}
\includegraphics[width=1.0\columnwidth]{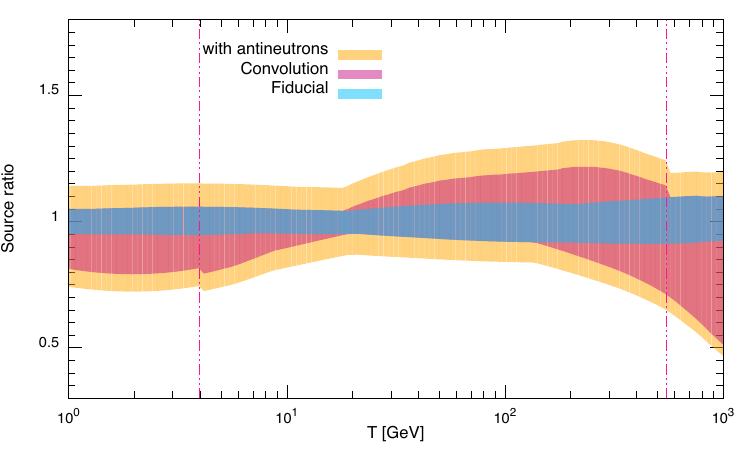} 
\end{center}
\caption{Estimate of the uncertainties in the antiproton source term from inelastic $pp$ scattering. The blue band 
indicates the 3$\sigma$ uncertainty band due to the global fit with Eq.\eqref{eq:NewParametrization2}, while the red band
corresponds to the convolution of the uncertainties brought by fits to the data with Eq.\eqref{eq:NewParametrization2},
Eq.\eqref{eq:NewParametrization3} and with the spline interpolation (see Fig.\ref{fig:figall}.).  The orange band takes into account the 
contribution from decays of antineutrons produced in the same reactions. Vertical bands as in Fig.\ref{fig:figall}. See text for details.}
\label{fig:sourcefinal}
\end{figure}

As a practical summary of our analysis, we report in Fig. \ref{fig:sourcefinal} an estimate for the 
uncertainties inherent to the production of antiprotons from inelastic $pp$ scatterings. 
The results are expressed as the ratio of the antiproton source term in Eq.\eqref{source} to a reference value. 
For the blue and the red bands, this reference value has been fixed to the source term obtained by setting
the $pp$ production cross section  to the best fit  to all the data obtained with Eq.~\eqref{eq:NewParametrization2} 
(parameters as in Table \ref{tab:all}). Outside the vertical bands---delimiting the energy range in which data are available---we 
extrapolate the production cross section by means of the same formula.

The blue band corresponds to considering parametrization \eqref{eq:NewParametrization2} alone. By simple inspection we can clearly see that the relevant uncertainty is maximally of the order of $10\%$. The red band is obtained by convoluting the uncertainty 
bands resulting from fits through Eqs.\eqref{eq:NewParametrization2} and
\eqref{eq:NewParametrization3} and (within the vertical bands) the spline interpolation.  This more conservative approach  
sizes the uncertainties from 20\% at the lowest energies to the extrapolated 50\% at 1 TeV. 

The most conservative estimate is shown by the orange band, where the additional uncertainty on the antineutron production has been taken into account. 
In this case, the normalization has been fixed to a source term 
in which the antineutrons produced in $pp$ scatterings contribute  with an energy-independent rescaling factor $\kappa = 1.3$ (w.r.t. 1).
The relevant uncertainty band has been derived by shifting the (red) previous convolution by an additional factor to account for 
the antineutron decay, $\kappa\simeq 1.3\pm 0.2$, as discussed in Sect. \ref{sec:therest}. 
The orange band indicates that the antiproton source term may 
vary by  $30\%$ at 1 GeV, and by up to more than $50\%$ at 1 TeV. In the energy range where scattering data are available 
(between 4 GeV  and 550 GeV), the uncertainty is of ${\cal{O}}(20-30\%)$ non-symmetrically around the reference source term, with the
ambiguity increasing with energy.

Finally, although we did not include an analysis of the uncertainties due to the contribution of nuclei to the yield, it is obvious that the total relative uncertainty cannot
be smaller than the one showed above.

\subsection{Conclusions}\label{sec:conclusions}

In this work we have performed a re-evaluation of the Lorentz-invariant distribution for inclusive antiproton production in 
$pp$ collisions in light of the recent results from the NA49 and BRAHMS experiments. We have combined these observations with older
measurements at different centre-of-mass energies in order to extract a reliable estimate both for the average
value of the cross section and for the current status of the corresponding uncertainties. 
Our main results have been presented in sections
\ref{sec:NA49}, \ref{sec:global} and \ref{sec:therest}
and summed up as handy fitting functions.

We have paid special attention in quantifying the extent to which the functional form we adopt introduces theoretical 
bias, by comparing our primary results to those obtained by using
different parametrizations, data-driven estimates or different subsets of the 
data. We are therefore confident that the uncertainties quoted throughout our paper are robust (cf also the discussion
in section \ref{sec:discussion}).

Our findings show that, despite the experimental progress in measuring
the inclusive antiproton production cross section, uncertainties
persist. In most of the well-constrained intermediate antiproton energy range (few GeV - few hundreds GeV), 
these uncertainties can be as low as $10 - 15\%$. 
At higher energies, we have shown that our knowledge is much worse, with extrapolations leading to
errors larger than $\sim 50\%$ at $1$ TeV.
 
A complete determination of the antiproton yield from $pp$ scattering must include the 
antineutron decay contribution, usually assumed to be identical to the antiproton one.
A recent measurement has reported a significant isospin dependence, which could amount 
to a slight increase of the antiproton source term and be accompanied by a non negligible uncertainty 
in the overall flux.

Last but not least, we remind that a significant contribution to the cosmic antiproton flux
is due to the reactions involving helium nuclei, both in the incoming radiation and in the ISM. The relevant 
cross sections have {\it never} been measured, and the corresponding antiproton source term can be estimated 
only by relying on some theoretical models and extrapolations, introducing an additional uncertainty which is hard to quantify.

We expect that for the few years to come, these uncertainties will continue imposing non-negligible limitations
on the interpretation of cosmic antiproton data, expected to be measured by AMS-02 with an unprecedented accuracy.
As a side-effect,  it appears unlikely that any definite conclusion for dark matter 
indirect detection could be drawn from a relatively featureless ``excess'' in the antiproton yield, expected to be be below a few tens of percent 
of the overall flux, unless perhaps correlated features are found in several different cosmic ray channels. At very low energies, the situation is further 
aggravated by the poor knowledge of astrophysical parameters, plus additional nuclear and particle physics uncertainties related e.g. to non-catastrophic
energy losses of antiprotons.

Despite current limitations, many of these sources of error are not irreducible, but could be addressed with dedicated experimental campaigns. As we stressed,
the key advances would come from planning experiments with an extended coverage in the $(p_T,x_R)$ plane, like NA49, with higher energies, 
with helium target/projectile, and by providing systematic analyses of anti-neutron yields. 
Another very useful enterprise  would be a re-analysis of existing
datasets with the aim of assessing the feed-down corrections associated to
different experiments. Indeed, the CR antiprotons must include  the
contributions from hyperon decays, while the current databases are not
uniform in that respect.

Cosmic ray antiprotons offer an important tool to address a number of astrophysical and astroparticle questions, as certified by the 
resources invested in balloon-borne and space-borne detectors in recent years. 
 We  believe that a nuclear and particle physics commitment to measure many of these missing ingredients should be identified  
 as a strategic task, to provide astroparticle missions with the crucial laboratory input to fully exploit their results.{}\\

\appendix

{\it Note Added:}  During the completion of this work, we became aware of a study by R. Kappl and M. W. Winkler, arXiv:1408.0299. On one side, it focuses  on 
NA49 data interpolation rather than performing a global dataset reanalysis with fitting formulae, like ours. 
On the other hand, it extends to propagated fluxes including other channels as well. As far as 
a comparison is possible, our results agree with theirs within quoted errors. 

\acknowledgments{We would like to thank H. G.Fischer for valuable information about NA49 experiment, D. Maurin for stimulating discussions and for careful reading of the manuscript, 
and R.H. Bernstein for spotting a typo in  Eq.~(\ref{eq:sigmainDuperray}) and another one in Table~(\ref{tab:duparry}) in an earlier version of this manuscript.
A.G. would like to acknowledge useful discussions with P. Aurenche on issues in hadron physics.
M.D.M. would like to thank Andrea Vittino for interesting and useful discussions on antiprotons secondary production uncertainties.
 At LAPTh, this activity was developed coherently with the research axes supported by the ANR Labex grant ENIGMASS. 
 F.D. further acknowledges support from AAP-``LeSAdHE'' of University of Savoie.}

\section{Useful kinematics} \label{sec:Kinematics}
In order to extract  the invariant
distribution of Eq.\eqref{eq:distributiongeneral}, we have transformed all data (when needed) in the CM frame, expressed them in the form
$(p_T, x_R, f(\sqrt{s}, p_T, x_R))$ and then performed our fit and interpolation as described in the main text. 
Here we collect for convenience some useful formulae by means of which these conversions can be done. 
In what follows, all kinematic variables carrying stars are defined
in the CM frame whereas those carrying $LAB$ superscripts are defined in the lab frame. 

Note that experiments \cite{Dekkers:1965zz,Allaby:1970jt,Antreasyan:1978cw} and \cite{Anticic:2009wd} are fixed-target experiments, whereas
\cite{Capiluppi:1974rt,Guettler:1976ce,Johnson:1977qx} and \cite{Arsene:2007jd} are colliding beam ones. With the exception of NA49 \cite{Anticic:2009wd}, fixed-target
experiments give their results in the LAB (or, equivalently in this case, target) frame. All other datasets are given in the CM frame.
\\
\\
The experiments \cite{Dekkers:1965zz} and \cite{Allaby:1970jt} do not provide results for the quantity defined in \eqref{eq:distributiongeneral},
 rather for $d^2 \sigma/d\Omega dp$ which can be recast as \cite{Byckling:1971vca}
\begin{equation}
\frac{d^2 \sigma}{d\Omega dp} = p_{\bar{p}}^{LAB,2} \frac{d^3\sigma}{dp^3} .
\end{equation}
This leads to
\begin{equation}
E_{\bar{p}}^{LAB} \frac{d^3\sigma}{dp_{\bar{p}}^3} = \frac{E_{\bar{p}}^{LAB}}{p_{\bar{p}}^{LAB,2}} \frac{d^2 \sigma}{d\Omega dp}
\label{eq:Allaby1970LorentzInv}
\end{equation}
which can be computed straightforwardly from the available data using $E_{\bar{p}}^{LAB} = \sqrt{p_{\bar{p}}^{LAB,2} + m_p^2}$.

The results are given as a function of the transverse momentum component $p_T$ 
and $\theta^{LAB}$, the $\bar{p}$ emission angle with respect to the incoming proton beam
in the lab frame.
Since $p_T$ remains unchanged in the two reference frames, one needs to compute the $x_R$ values which the measurements correspond to.
We have that
\begin{align}
\label{eq:Epbartransf}
E_{\bar{p}}^* = \gamma^{CM} (E_{\bar{p}}^{LAB} - v^{CM} p_{\bar{p}}^{LAB} \cos\theta^{LAB})
\end{align}
where $v^{CM}$ and $\gamma^{CM}$ are the velocity of the CM reference frame with respect to the lab one and the corresponding Lorentz boost.
These are given by
\begin{align}
\label{eq:cmeq}
v^{CM} & = p_p^{LAB}/(E_p^{LAB} + m_p)  \\ \nonumber
& = p_p^{LAB}/(\sqrt{p_p^{LAB\,^2} + m_p^2} + m_p) \\ \nonumber
\gamma^{CM} & = (E_p^{LAB} + m_p)/\sqrt{s}\,.
\end{align}
Then, direct use of the definition \eqref{eq:xRdef} allows us to compute $x_R$.
\\
\\
Experiments \cite{Capiluppi:1974rt,Guettler:1976ce} and \cite{Anticic:2009wd} provide results for $f(p+p \rightarrow \bar{p}+X)$ 
in the CM frame, as a function of $p_T$ and the alternative Feynman scaling variable $x_F$ defined as 
\begin{equation}
x_F = \frac{2 p_L^*}{\sqrt{s}} \simeq \frac{p_L^*}{p_{L,max}^*}\,,
\label{eq:xFdef}
\end{equation}
where $p_L^*$ is the antiproton longitudinal momentum and $p_{L,max}^*$ is its maximum possible longitudinal momentum. It is thus necessary
to find the correspondence between the $x_R$ and $x_F$ scaling variables.
First, from $p_L^{*2} = p^{*2} - p_T^{*2}$ and $p^2 = E^2 - m^2$ we get 
\begin{equation}
E_{\bar{p}}^{*2} = p_L^{*2} + m_p^2 + p_T^{*2} .
\label{eq:Epbarstar}
\end{equation}
whereas $E^{*}_{\bar{p}\,\rm{max}}$ is given by \eqref{sqrtslab}. Direct use of \eqref{eq:xFdef} and \eqref{eq:xRdef} then gives
\begin{equation}
x_R = \frac{\sqrt{x_F^2 (s/4) + m_p^2 + p_T^2}}{E_{\bar{p},\mathrm{max}}^*}\,.
\label{eq:xFxRexact}
\end{equation}
\\
\\
Reference \cite{Antreasyan:1978cw} provides results for $f(p+p \rightarrow \bar{p}+X)$ in the CM frame as a function of $p_T$ 
and $\theta^{LAB}$, also providing the values $\theta^{*}$ of the angle in the CM frame.
The $p_T$ values remain unchanged in the CM frame. So all that is left is to calculate $x_R$. 
Since $p_{\bar{p}}^* = p_T/\sin\theta^*$, by using $E = \sqrt{\vec{p}^2 + m_p^2}$ and equation \eqref{eq:estarmax}, 
we can calculate $x_R$ through \eqref{eq:xRdef}.
\\
\\
The BRAHMS experiment \cite{Arsene:2007jd} chooses to give its results for the invariant cross section
\begin{equation}
\frac{1}{2\pi p_T} \frac{d^2\sigma}{dp_T \, dy} = E_{\bar{p}} \frac{d^3\sigma}{dp_{\bar{p}}^3}
\end{equation}
as a function of the antiproton transverse momentum $p_T$ and the antiproton rapidity in the CM frame, $y^*$, defined as
\begin{equation}
y^* = \frac{1}{2} \ln \left( \frac{E^* + p_L^*}{E^* - p_L^*}\right) .
\label{eq:RapidityDef}
\end{equation}
From this definition and $E_{\bar{p}}^{*2} = p_L^{*2} + p_T^2 + m_{\bar{p}}^2$ we get 
\begin{equation}
E_{\bar{p}}^{*} = \sqrt{m_{\bar{p}}^2 + p_T^2} \cosh y . 
\end{equation}
Then, $x_R$ can be calculated by means of this relation and \eqref{eq:xRdef}.
\\
\\
 It is maybe useful to report the computation of the maximal antiproton energy $E^{*}_{\bar{p}\,\rm{max}}$ introduced in \ref{sqrtslab}.
For a general inclusive reaction $a+b \rightarrow c+X$ in the CM frame we have
\begin{align}
\sqrt{s} = E_c^* + E_X^* \ .
\end{align}
By replacing
$E_c^*$, $E_X^*$ through $E = \sqrt{p^2 + m^2}$, squaring the resulting relation, reintroducing the energies in the crossed terms, eliminating $E_X$ in favour of $\sqrt{s}$ and $E_c^*$, using $p^{*2} - E^{*2} = -m^2$ to eliminate $p^{*2}$ and solving for $E_c^*$ we get
\begin{equation}
E_c^* = \frac{s + m_c^2 - m_X^2}{2 \sqrt{s}}\,,
\label{eq:ecm}
\end{equation}
where we introduced for compactness a slight abuse of notation for the $X$ system by assigning 
it a mass variable $m_X$. In reality $m_X$ simply refers to internal energy of the $X$ system.

Now, for $s$ and $m_p$ fixed, we see that $E_{\bar{p},\mathrm{max}}^*$ is obtained through \eqref{eq:ecm} once the energy $m_X$
of the $X$ system becomes minimal. Conservation of baryon number fixes $m_{X,{\rm  min}} = 3 m_p$ (production of 3 additional protons at rest). 
So we get
\begin{equation}
E_{\bar{p},\mathrm{max}}^* = \frac{s - 8 m_p^2}{2 \sqrt{s}}\,.
\label{eq:estarmax}
\end{equation}

Finally we derive the maximum antiproton angle with respect to the incoming proton in the laboratory frame $\theta^{LAB}_{max}$. This quantity can be derived using the condition $x_{R}\leq 1$ and the definition of $x_{R}$ in Eq.\eqref{eq:xRdef}. Then writing $E_{\bar{p}}^{*}$ in the laboratory frame (see Eq.\eqref{eq:Epbartransf}) the condition $x_{R}\leq 1$ can be written as
\begin{equation}
x_{R} = \frac{\gamma^{CM} 2 \sqrt{s} (E_{\bar{p}}^{LAB} - v^{CM} p_{\bar{p}}^{LAB} \cos\theta^{LAB})}{s - 8 m_p^2} \leq 1\,.
\label{eq:thetamax}
\end{equation}
By using the definition of $v^{CM}$ and $\gamma^{CM}$ given in Eq.\eqref{eq:cmeq} we get
\begin{equation}
\cos \theta^{LAB}_{max} =  \frac{E_{\bar{p}}^{LAB} (E_{p}^{LAB} + m_p) - m_p(E_{p}^{LAB} - 3m_p)}{\sqrt{{E_{\bar{p}}^{LAB}}^2-m^2_p} \sqrt{{E_{p}^{LAB}}^2-m^2_p}}  \,.
\label{eq:thetamaxdef}
\end{equation}

\section{Inelastic cross section parameterization} \label{sec:sigmainel}

The inelastic proton cross section is defined as the difference between the total $pp$ scattering cross section 
$\mathrm{\sigma_{tot}^{pp}}$ and its elastic counterpart $\mathrm{\sigma_{el}^{pp}}$
\begin{equation}
\mathrm{\sigma_{in}^{pp}} = \mathrm{\sigma_{tot}^{pp}} - \mathrm{\sigma_{el}^{pp}} .
\label{eq:sigmappineldef}
\end{equation}

In order to estimate $\mathrm{\sigma_{in}^{pp}}$ for our energy region of interest, we employ the
experimental data provided by the Particle Data Group (PDG) on the total and elastic $pp$ cross sections \citep{PDGcs2013}. We fit this data
by means of the highest-ranking parametrization of the total proton cross section suggested by the PDG itself, which reads
\begin{equation}
\mathrm{\sigma_{tot}^{pp}} = Z^{pp} + B^{pp}\log^2(s/s_M) + Y_1^{pp}(s_M/s)^{\eta_1} - Y_2^{pp}(s_M/s)^{\eta_2}
\label{eq:sigmappinelparam}
\end{equation}
where $B^{pp} = \pi (\hbar c)^2/M^2$, $s_M = (2 m_p + M)^2$, all energies are given in GeV and all cross sections in mb. 

Although
this parametrization is given for the total proton cross section, noticing the resemblance in the $\sqrt{s}$ dependence of both the total
and the elastic cross section, we employ the same functional form in order to fit the data on both $\mathrm{\sigma_{tot}^{pp}}$
and $\mathrm{\sigma_{el}^{pp}}$.
In both cases, we include in the fitting procedure only data referring to $\sqrt{s} > 5$ GeV. 
Moreover, given the large amount of relevant data available for low $\sqrt{s}$ values
as opposed to the relative scarcity of measurements 
for $\sqrt{s} > 80$ GeV, we include the highest-energy data contained in the sample in order to capture the cross section behaviour 
over our entire energy range of interest in a reliable manner.

The fitting parameters for both $\mathrm{\sigma_{tot}^{pp}}$ and $\mathrm{\sigma_{el}^{pp}}$ are shown in 
Table \ref{tab:protonCSfit}. The resulting functions provide very good fits to the available data, with a $\chi^2$ per degree of
freedom of $0.78$ for the total cross section and $1.5$ for the elastic case, where statistical and systematic uncertainties have been added
in quadrature. With these two parametrizations at hand, we then compute the inelastic cross section through its definition
\eqref{eq:sigmappineldef}. We have verified that, modulo small differences due to the specific goals of our analysis described before,  
our results are in very good agreement not only with the PDG results 
on $\mathrm{\sigma_{tot}^{pp}}$ but also with the function for $\mathrm{\sigma_{el}^{pp}}$ quoted by all LHC experiments
(see e.g. \cite{Antchev:2013iaa,Antchev:2013paa}) as well as with the  results on both 
$\mathrm{\sigma_{tot}^{pp}}$ and $\mathrm{\sigma_{el}^{pp}}$ by Block and Halzen \cite{Block:2012ym}.

\renewcommand{\arraystretch}{1.1}
\begin{table}
\begin{center} 
\vspace{1cm}
\begin{tabular}{|c|c|c|} 
\hline \hline 
Parameter & Total & Elastic                  \\ 
\hline 
$M$        &  2.06  &  3.06                  \\ 
$Z^{pp}$   &  33.44 &  144.98                \\ 
$Y_1^{pp}$ &  13.53 &  2.64                  \\ 
$Y_2^{pp}$ &  6.38  &  137.27                \\
$\eta_1$   &  0.324 &  1.57                  \\
$\eta_2$   &  0.324 &  -4.65$\times 10^{-3}$ \\
\hline \hline
\end{tabular} \caption{Fit results for the total and elastic proton scattering cross sections according to Eq.\eqref{eq:sigmappinelparam}.
\label{tab:protonCSfit}}
\end{center} 
\end{table}
\renewcommand{\arraystretch}{}

\bibliography{sigmapbardraft}
\end{document}